\documentstyle[prl,preprint,eqsecnum,aps,epsf]{revtex}
\begin{document}

\title{Low-Energy Absorption Cross Section for massive scalar and Dirac  
fermion by $(4+n)$-dimensional Schwarzschild Black Hole} 
\author{Eylee Jung\footnote{Email:eylee@kyungnam.ac.kr}, 
SungHoon Kim\footnote{Email:shoon@kyungnam.ac.kr} and
D. K. Park\footnote{Email:dkpark@hep.kyungnam.ac.kr 
}, }
\address{Department of Physics, Kyungnam University, Masan, 631-701, Korea}

\maketitle

\maketitle
\begin{abstract}
Motivated by the brane-world scenarios, we study the absorption
problem when the spacetime background is $(4+n)$-dimensional Schwarzschild 
black hole. We compute the low-energy absorption cross sections for the 
brane-localized massive scalar, brane-localized massive Dirac fermion, and 
massive bulk scalar. For the case of brane-localized massive Dirac fermion 
we introduce the particle's spin in the traditional Dirac form without 
invoking the Newman-Penrose method. Our direct introduction of spin enables 
us to 
compute contributions to the $j$th-level partial absorption cross section 
from orbital angular momenta $\ell = j \pm 1/2$. It is shown that the
contribution from the low $\ell$-level
is larger than that from the high $\ell$-level 
in the massive case. In the massless case these two contributions 
are exactly same with each other. The ratio of low-energy absorption cross
sections for Dirac fermion and for scalar is dependent on the number of extra
dimensions as $2^{(n-3)/ (n+1)}$. Thus the ratio factor $1/8$ is recovered 
when $n=0$ as Unruh found long ago. 
The physical importance of this ratio factor is discussed in the context of
the brane-world scenario.
For the case of bulk scalar our 
low-energy absorption cross section for S-wave is exactly same with area of the 
horizon hypersurface in the massless limt, which is an higher-dimensional
generaliztion of universality. Our results for all cases turn out to
have correct
massless and $4d$ limits.

\end{abstract}

\newpage
\section{Introduction}
The ``greybody factor'' $\Gamma_{\ell}(\omega)$ is an important quantity to 
understand the absorption and emission phenomena of a black hole. It is 
this factor which makes a black hole to be different from a black body.
The physical origin of this factor is an effective potential barrier 
generated by a black hole spacetime. For example, the potential for the 
massless scalar generated by the $4d$ Schwarzschild spacetime is 
\begin{equation}
\label{effective1}
V_{\mbox{eff}}(r_{\ast}) = 
\left(1 - \frac{r_H}{r}\right) 
\left(\frac{r_H}{r^3} + \frac{\ell (\ell + 1)}{r^2}\right)
\end{equation}
when the wave equation is expressed in terms of the ``tortoise'' coordinate
$r_{\ast} = r + r_H \ln (r / r_H - 1)$, where $r_H$ is an horizon radius and
$\ell$ is an angular momentum of the scalar. This potential generally 
backscatters a part of the outgoing radiation quantum machanically, which 
results in the frequency-dependent greybody factor $\Gamma_{\ell}(\omega)$.
This factor , therefore, appears in the black hole's thermal radiation
formula
\begin{equation}
\label{radiation1}
\frac{dH}{d \omega} = \sum_{\ell} 
\frac{\Gamma_{\ell}(\omega)}{e^{\omega / T_{BH}} - 1}
\frac{(2\ell + 1) \omega}{\pi}
\end{equation}
where $T_{BH}$ is an Hawking temperature. Eq.(\ref{radiation1}) shows how the
greybody factor and the Planck factor play important roles in the emission 
rate of energy. 

In addition to the radiation formula (\ref{radiation1}) the 
factor $\Gamma_{\ell}(\omega)$ is important to compute the partial 
absorption cross section $\sigma_{\ell}(\omega)$ of the black hole. The
explicit relation\footnote{The relation between 
$\sigma_{\ell}(\omega)$ and $\Gamma_{\ell}(\omega)$ in higher dimensions is 
given in Eq.(\ref{optical}).} 
 between $\sigma_{\ell}(\omega)$ and $\Gamma_{\ell}(\omega)$
for the $4d$ massive scalar is 
\begin{equation}
\label{relation1}
\sigma_{\ell}(\omega) = \frac{\pi}{\omega^2 v^2} (2\ell + 1)
\Gamma_{\ell}(\omega)
\end{equation}
and $v = \sqrt{1 - m^2 / \omega^2}$ and $m$ is particle's mass. Thus one can 
compute $\sigma_{\ell}$ from $\Gamma_{\ell}$ straightforwardly or 
{\it vice versa}. 

Many computational techniques for calculation  of the absorption cross section 
in the various
$4d$ black hole were developed long 
ago\cite{teuk72,sta73,cart74,ford75,page76,unruh76,sanc78}. The main streams of 
this procedure are to derive the solutions of the given wave equation in the 
near-horizon and asymptotic regimes separately and to match them in the 
appropriate intermediate place. Following this procedure Unruh\cite{unruh76}
computed the low energy absorption cross section for the massive scalar and
Dirac fermion in the $4d$ Schwarzschild background. It is instructive to 
introduce an explicit expression of Ref.\cite{unruh76} for the massive 
scalar:
\begin{eqnarray}
\label{unruh1}
& &(\sigma_{\ell})_{unruh} = \frac{\pi}{k^2 v^2} (2 \ell + 1) T_{\ell}
                                                          \\   \nonumber
& &T_{\ell} = \frac{\pi (\ell !)^4 2^{2 \ell + 2} (1 + v^2) k^{2 \ell + 3}
                  v^{2 \ell}}
                 {(2 \ell!)^4 (2 \ell + 1)^2
                  [1 - \exp\left\{-\pi k (1 + v^2) / v \right\}]}
            \prod_{s = 1}^{\ell}
            \left[s^2 + \left(\frac{k (1 + v^2)}{2 v} \right)^2 \right].
\end{eqnarray}

The most interesting one Eq.(\ref{unruh1}) suggests is the fact that the 
low-energy absorption cross section for S-wave is equal to the horizon 
area in the massless limit. 
This is an universal property for the minimally coupled massless scalar.
This universality indicates that the low-energy cross section encodes 
information on the near-horizon structure of black hole. Another interesting 
result of Ref.\cite{unruh76} is the fact that the low-energy absorption 
cross section for Dirac fermion with mass $\mu$ is exactly $1/8$ of that 
for scalar with mass $m$ if $\mu = m$. It is still unclear at least for us 
the physical origin of this ratio factor. 

The universality is re-examined in the arbitrary dimensional spherically
symmetric black hole\cite{das97}. Ref.\cite{das97} has shown that the 
low-energy absorption cross sections for scalar is equal to the horizon
area while that for spin-$1/2$ particle is an area measured in a flat 
spatial metric conformally related to the true metric. The universality 
for the minimally coupled massless scalar is extended to the 
$p$-brane-like objects\cite{emp98} and its generalization to the massive 
scalar is discussed recently\cite{park00,jung03,jung04}. In particular,
it is shown in Ref.\cite{jung03} that the mass-dependence of the absorption 
cross section is very sensitive to the near-horizon structure of spacetime.

The computation of the low-energy absorption cross section is also important 
subject in the context of string theories and brane-world scenario. In 
string theories the black hole is effectively represented by a collective
states\cite{mal98-1} and the relevant low energy excitations of this effective
description are the right- and left-moving modes of the string. The computation
of the low-energy absorption cross section with this picture suggests that 
the effective string theory for the black hole is ``heterotic'', {\it i.e.}
the right-moving sector has both fermionic and bosonic degrees of freedom, 
while the left-moving sector has only bosonic degree of 
freedom\cite{mal96-1,cve98-1}. 

In the context of the brane-world scenario the most remarkable fact is that
the fundamental Planck mass can be low as a TeV scale. This TeV-scale
gravity is realized by making use of the large extra 
dimensions\cite{ark98-1,anto98} or warped extra dimensions\cite{rs99-1}. 
One of the striking consequences arising due to the TeV-scale gravity is that
the future high-energy colliders such as the CERN Large Hadron Collider 
can be a black hole factory\cite{gidd02-1,dimo01-1,eard02-1}. If the 
black holes can be really produced in the future collider, one can examine 
the quantum gravity effects of black hole in the laboratory such as 
Hawking radiation and/or information loss problem\cite{hawk76,horo04}. Thus, 
it is important to investigate the absorption and emission problems in this 
context. Recently works along this direction 
were done\cite{kanti02-1,kanti03-1,harris03-1}. In Ref.\cite{kanti02-1} the 
low-energy absorption problems for massless bulk and brane-localized scalar
are examined. In Ref.\cite{kanti03-1} same problems for brane-localized 
massless particle with spin $1/2$ and $1$ are examined. The absorption and 
emission problems for the brane-localized particles in the full range of 
energy are numerically studied in Ref.\cite{harris03-1}.

In this paper we would like to extend Ref.\cite{kanti02-1,kanti03-1} by 
considering the massive spin-$0$ and spin-$1/2$ particles in the 
$(4+n)$-dimensional Schwarzschild background. Although the spin
can be introduced in the curved spacetime with ease by making use of the 
Newman-Penrose formalism\cite{newman62}, we will introduce it in the more
traditional Dirac form. In section II we will compute the low-energy 
absorption cross section for the brane-localized massive scalar.
It is shown in this section  that our final result for the low-energy 
absorption cross section has a correct massless
limit. However, the $4d$ limit of it is not exactly same with 
Eq.(\ref{unruh1}), but coincides with the low-energy expansion of 
Eq.(\ref{unruh1}). The reason for this is explained in the appendix.
In section III the low-energy absorption cross section for the massive 
Dirac fermion is computed by solving Dirac equation. 
The final form of the low-energy absorption cross section 
$\sigma_j(\omega)$, where $j$ is a total angular momentum, is just sum of 
two contributions from orbital angular momenta $\ell = j \pm 1/2$, {\it i.e.}
$\sigma_j(\omega) = \sigma_{j,\ell=j+1/2}(\omega) + \sigma_{j,\ell=j-1/2}(\omega)$. 
While the introduction of spin via usual Newman-Penrose formalism generally
does not allow to compute each contribution, our introduction of spin enables
us to compute $\sigma_{j,\ell=j+1/2}(\omega)$ and 
$\sigma_{j,\ell=j-1/2}(\omega)$ separately. It is shown in this section that 
the contribution from lower $\ell$-state is larger than that from higher
$\ell$-state in the massive case. However, these two contributions to 
$\sigma_j(\omega)$ are exactly same in the massless limit. It is also
shown that the ratio of the absorption cross section for massive scalar and
massive Dirac fermion is $2^{(n-3)/(n+1)}$, which gives a factor $1/8$ 
when $n=0$ as Unruh found.
In section IV
the low-energy absorption cross section for bulk scalar is computed. 
It is shown that the massless and S-wave limit of our result for the absorption
cross section is same with the area of the horizon hypersurface. Thus, the 
universality for S-wave is preserved in the higher-dimensional theories.
In section V a brief conclusion is given.

\section{Low-Energy Absorption Cross Section for Brane-Localized Scalar}
The various higher-dimensional black hole solutions of the Einstein equation
were discussed in detail in Ref.\cite{myers86}. The explicit expression of the
$(4+n)$-dimensional Schwarzschild solution in terms of the usual Schwarzschild
coordinates is in the following:
\begin{equation}
\label{schwarz1}
ds^2 = -h(r) dt^2 + h(r)^{-1} dr^2 + r^2 d\Omega_{n+2}^2
\end{equation}
where
\begin{equation}
\label{horizonp}
h(r) = 1 - \left(\frac{r_H}{r}\right)^{n+1}
\end{equation}
and the angular part is given by
\begin{equation}
\label{angle-part}
d\Omega_{n+2}^2 = d\theta_{n+1}^2 + 
\sin^2 \theta_{n+1} \Bigg( d\theta_{n}^2 + \sin^2 \theta_n \bigg(
\cdots + \sin^2 \theta_2 \left( d\theta_1^2 + \sin^2 \theta_1 d\phi^2
           \right) \cdots \bigg) \Bigg).
\end{equation}
The horizon radius $r_H$ is related to the black hole mass $M$ as following:
\begin{equation}
\label{relation2}
r_H^{n+1} = \frac{8 \Gamma \left(\frac{n+3}{2}\right)}
                 {(n+2) \pi^{\frac{n+1}{2}}}
            \frac{M}{M_{\ast}^{n+2}}
\end{equation}
where $M_{\ast} = G^{-1/(n+2)}$ is a $(4+n)$-dimensional Planck mass and 
$G$ is a Newton constant.

Since we are interested in the scalar localized on the brane in this section,
we assume the scalar field $\Phi$ is a function of only $t$, $r$, $\theta_1$, 
and $\phi$. Thus, letting 
$\Phi=e^{-i \omega t} R_{\omega \ell} (r) Y_{\ell}(\theta_1, \phi)$ yields a 
following radial equation
\begin{equation}
\label{radial1}
\frac{h(r)}{r^2} \frac{d}{d r} \left[h(r) r^2 \frac{d R}{d r}\right]
+ \left[\omega^2 - h(r) \left\{m^2 + \frac{\ell (\ell + 1)}{r^2} \right\}
                                        \right] R = 0.
\end{equation}

For the quantum-mechanical interpretation we can change Eq.(\ref{radial1}) into
the following Schr\"{o}dinger-like equation
\begin{equation}
\label{schro1}
-\frac{1}{2 M_{eff}} \frac{d^2 \psi_{\ell}}{d r_{\ast}^2}
+ V_{eff}(r_{\ast}) \psi_{\ell} = \omega^2 v^2 \psi_{\ell}
\end{equation}
where
\begin{eqnarray}
\label{bozo1}
& &R = r^{\frac{n}{2}} \psi_{\ell}, \hspace{2.0cm}
r_{\ast} = r_{H} \ln h
                                    \\   \nonumber
& &M_{eff}^{-1} = 2 (n+1)^2 (1 - h)^{\frac{2n + 4}{n + 1}}
                                    \\    \nonumber
& &V_{eff}(r_{\ast}) = h \frac{\ell (\ell + 1) - 
                               \frac{n}{2} \left[(n+1) -\frac{n}{2} h \right]}
                              {r^2}
                      - (1 - h) m^2.
\end{eqnarray}
Unlike the usual $4d$ black hole case the Schr\"odinger-like equation
(\ref{schro1}) involves not only the effective potential but also the 
effective position-dependent mass. Thus, it seems to be difficult to get a
direct quantum-mechanical interpretation from (\ref{schro1}). However, one 
can roughly estimate the effect of extra dimension in the absorption 
cross section. For simplicity let us consider the massless limit. The 
effective potential as a function of $r_{\ast}$ is plotted in Fig. 1 at 
fixed $\ell$. Fig. 1 indicates that the potential barrier height seems to 
increase with increasing the number of extra dimensions. Since the 
greybody factor usually decreases when potential barrier becomes higher and
the absorption cross section is proportional to this factor, one can 
conjecture that the existence of the extra dimensions may reduces the 
absorption cross section, which explains Fig. 1 of Ref.\cite{harris03-1}.
The detailed analysis of the Schr\"odinger-like equation (\ref{schro1}) will
be discussed elsewhere.

In Ref.\cite{unruh76} Unruh derived the low-energy absorption cross section
by solving the wave equation, {\it i.e.} the corresponding equation of 
Eq.(\ref{radial1}) in $4d$ Schwarzschild black hole, in the near-horizon, 
asymptotic and intermediate regions separately and matching the 
near-horizon and asymptotic solutions via the solution in the intermediate
region. One may follow this procedure, but it seems to be impossible unlike
$4d$ case to derive a solution in the intermediate region analytically. Thus,
we adopt an alternative method introduced by Maldacena and Strominger in 
Ref.\cite{mal96-1}. As will be shown in the next section, however, this 
alternative method does not work when we discuss the absorption problem 
for Dirac fermion. In that case we will adopt the Unruh's original 
method.

Changing a variable makes Eq.(\ref{radial1}) to be 
\begin{equation}
\label{var1}
h (1-h) \frac{d^2 R}{d h^2} + 
\left(1 - \frac{2n+1}{n+1} h \right) \frac{d R}{d h}
+ \frac{1}{(n+1)^2}
\left[ \frac{(\omega r)^2}{h (1 - h)} - 
       \frac{(m r)^2 + \ell (\ell + 1)}{1 - h} \right] R = 0.
\end{equation}
In the near-horizon region, {\it i.e.} $r \sim r_H$, Eq.(\ref{var1}) is solved
in terms of the hypergeometric function as follows:
\begin{eqnarray}
\label{nearh1}
R_{NH}(r)&=& h^{\alpha} (1 - h)^{\beta}
\Bigg[A_- 
F\left(\alpha + \beta + \frac{n}{1 + n}, \alpha + \beta ; 1 + 2 \alpha ; h
                                                                      \right)
                                        \\   \nonumber
& & \hspace{3.0cm} + A_+ h^{-2 \alpha}
F\left(-\alpha + \beta + \frac{n}{1 + n}, -\alpha + \beta ; 1 - 2 \alpha ; h
                                                                      \right) 
                                                                 \Bigg]
\end{eqnarray} 
where 
\begin{eqnarray}
\label{bozo2}
\alpha&=& - \frac{i \omega r_H}{n + 1}        \\   \nonumber
\beta&=& \frac{1}{2(n+1)}
\left[1 - \sqrt{(2 \ell + 1)^2 - 4 \omega^2 v^2 r_H^2} \right].
\end{eqnarray}
Since $h \sim e^{r_{\ast} / r_H}$ in terms of tortoise coordinate 
introduced in Eq.(\ref{bozo1}) and only outgoing wave is valid in the 
near-horizon region, we should impose $A_+ = 0$ and then $R_{NH}(r)$ reduces
to 
\begin{equation}
\label{nearh2}
R_{NH}(r) \sim A_- e^{-i \frac{\omega r_{\ast}}{n + 1}}.
\end{equation}

Now, let us solve Eq.(\ref{radial1}) in the asymptotic region.
Putting $h \sim 1$ makes Eq.(\ref{radial1}) in the following simple form; 
\begin{equation}
\label{asymp1}
\frac{d^2 R_{FF}}{d r^2} + \frac{2}{r} \frac{d R_{FF}}{d r}
+ \left[\omega^2 v^2 - \frac{\ell (\ell + 1)}{r^2} \right] R_{FF} = 0
\end{equation}
which yields an solution in terms of the usual Bessel function as 
following:
\begin{equation}
\label{asymp2}
R_{FF} = \frac{1}{\sqrt{r}}
\left[ B_+ J_{\ell + \frac{1}{2}} (\omega v r) + 
       B_- Y_{\ell + \frac{1}{2}} (\omega v r)  \right].
\end{equation}

To match the near-horizon solution $R_{NH}(r)$ in Eq.(\ref{nearh1}) with 
$A_+ = 0$ and the far-field $R_{FF}(r)$ in Eq.(\ref{asymp2}), we change the 
near-horizon solution as following:
\begin{eqnarray}
\label{transf1}
& &R_{NH}(r)= A_- h^{\alpha} (1 - h)^{\beta}
\Bigg[\frac{\Gamma \left(1 + 2 \alpha \right) 
             \Gamma \left(-2 \beta + \frac{1}{n+1} \right)}
           {\Gamma \left(\alpha - \beta + \frac{1}{n+1} \right)
            \Gamma \left(1 + \alpha - \beta \right)}   
                                                     \\ \nonumber
& & \hspace{7.0cm} \times
     F \left( \alpha + \beta + \frac{n}{n+1}, \alpha + \beta; 
             2\beta + \frac{n}{n+1}; 1-h \right)
                                                   \\   \nonumber
& & 
\hspace{5.0cm}
+ (1-h)^{-2 \beta + \frac{1}{n+1}}
  \frac{\Gamma \left(1 + 2 \alpha \right) 
        \Gamma \left(2\beta - \frac{1}{n+1} \right)}
       {\Gamma \left(\alpha + \beta + \frac{n}{n+1} \right)
        \Gamma \left(\alpha + \beta \right)}
                                                  \\  \nonumber
& & \hspace{7.0cm} \times
     F \left(\alpha - \beta + \frac{1}{n+1}, 1 + \alpha - \beta;
             -2\beta + \frac{n+2}{n+1}; 1-h \right)      \Bigg].
\end{eqnarray}
Taking $R_{NH}(r)$ in Eq.(\ref{transf1}) to $r \rightarrow \infty$ and 
$R_{FF}(r)$ in Eq.(\ref{asymp2}) to $r \rightarrow 0$ naturally yields 
two relations, one between $A_-$ and $B_+$ and the other between 
$A_-$ and $B_-$. Then removing $A_-$ yields
\begin{equation}
\label{bbb1}
B \equiv \frac{B_+}{B_-} = 
-\left( \frac{2}{\omega v r_H} \right)^{2\ell + 1}
\frac{\left(\ell + \frac{1}{2}\right) \Gamma^2 \left(\ell + \frac{1}{2} \right)
      \Gamma \left(\frac{1}{n+1} - 2\beta \right)
      \Gamma \left(\alpha + \beta + \frac{n}{n + 1} \right)
      \Gamma \left(\alpha + \beta \right) }
     {\pi \Gamma \left(1 + \alpha - \beta \right)
      \Gamma \left(\alpha - \beta + \frac{1}{n+1} \right)
      \Gamma \left(2 \beta - \frac{1}{n+1} \right)}.
\end{equation}

Computing the flux in the asymptotic region, one can calculate the 
greybody factor whose expression is 
\begin{equation}
\label{grey1}
\Gamma_{\ell}(\omega) = 1 - \Bigg| \frac{B - i}{B + i} \Bigg|^2
= \frac{2 i (B^{\ast} - B)}{|B|^2 + i (B^{\ast} - B) + 1}
\approx \frac{2 i (B^{\ast} - B)}{|B|^2}.
\end{equation}
The last approximation in Eq.(\ref{grey1}) comes from the low-energy 
approximation, {\it i.e.} $\omega << 1$. Inserting Eq.(\ref{bbb1}) into 
Eq.(\ref{grey1}) makes the greybody factor to be 
\begin{equation}
\label{grey2}
\Gamma_{\ell}(\omega) = \frac{16 \pi}{(n+1)^2 v}
\left(\frac{\omega v r_H}{2}\right)^{2 \ell + 2}
\frac{\Gamma^2 \left(1 + \frac{\ell}{n+1} \right)
      \Gamma^2 \left( \frac{\ell + 1}{n + 1} \right)}
     {\Gamma^2 \left(\ell + \frac{1}{2} \right)
      \Gamma^2 \left(1 + \frac{2 \ell + 1}{n + 1} \right)}.
\end{equation}
Then the partial absorption cross section $\sigma_{\ell}$ can be read
straightforwardly from Eq.(\ref{grey2});
\begin{equation}
\label{partial1}
\sigma_{\ell} \equiv \frac{\pi (2 \ell + 1)}{\omega^2 v^2} 
                        \Gamma_{\ell}(\omega)
= \frac{16 \pi^2 (2 \ell + 1)}{(n+1)^2 \omega^2 v^3}
\left(\frac{\omega v r_H}{2}\right)^{2 \ell + 2}
\frac{\Gamma^2 \left(1 + \frac{\ell}{n+1} \right)
      \Gamma^2 \left( \frac{\ell + 1}{n + 1} \right)}
     {\Gamma^2 \left(\ell + \frac{1}{2} \right)
      \Gamma^2 \left(1 + \frac{2 \ell + 1}{n + 1} \right)}.
\end{equation}

It is interesting to consider several limiting cases. Firstly, let us consider
the massless limit, {\it i.e.} $v \approx 1$. Then the partial absorption
cross section reduces to 
\begin{equation}
\label{nomass1}
\sigma_{\ell}^{m=0} = \frac{16 \pi^2 (2 \ell + 1)}{(n+1)^2 \omega^2}
\left(\frac{\omega r_H}{2}\right)^{2 \ell + 2}
\frac{\Gamma^2 \left(1 + \frac{\ell}{n+1} \right)
      \Gamma^2 \left( \frac{\ell + 1}{n + 1} \right)}
     {\Gamma^2 \left(\ell + \frac{1}{2} \right)
      \Gamma^2 \left(1 + \frac{2 \ell + 1}{n + 1} \right)}
\end{equation}
which is exactly same with the result of Ref.\cite{kanti02-1}. In 
$n \rightarrow 0$ limit Eq.(\ref{partial1}) becomes 
\begin{equation}
\label{noextra1}
\sigma_{\ell}^{n=0} = \frac{\pi (\ell !)^6 2^{2 \ell + 2} \omega^{2 \ell} 
                            v^{2 \ell - 1} }
                           {(2 \ell !)^4 (2 \ell + 1)}   r_H^{2 \ell + 2}
\end{equation}
which coincides with Eq.(\ref{unruh1}) if we expands Eq.(\ref{unruh1}) 
by making use of $\omega \rightarrow 0$ limit. Thus our result (\ref{partial1}) 
recovers not only the massless limit but also $4d$ limit.

At this stage one may question why our $4d$ limit is not exactly 
Eq.(\ref{unruh1}) but low-energy expansion of it. In fact, Unruh derived 
Eq.(\ref{unruh1}) by expressing the asymptotic solution in terms of the 
Coulomb wave functions. The multiplicative factor and exponential factor
in Eq.(\ref{unruh1}) are results of these Coulomb wave functions. In Appendix,
however, we will show that the expression of asymptotic solution in terms
of the Coulomb wave functions is impossible except $n=0$. This is why 
the $4d$ limit of our result gives the low-energy expansion of 
Eq.(\ref{unruh1}).

\section{Low-Energy Absorption Cross Section for Dirac Fermion}
In this section we will compute the low-energy absorption cross section 
for the massive Dirac fermion in the Schwarzschild background defined on the
bulk. As commented earlier, spin of the particle can be introduced more 
conveniently by exploiting the Newman-Penrose formalism. However, it is 
more straightforward to introduce it in the traditional Dirac form, which 
we will follow in this section. As will be shown shortly, furthermore, our
introduction of spin enables us to compute each contribution to the low-energy
partial absorption cross section. The usual Newman-Penrose formalism does not
provide this information. It usually gives the total sum of it. Thus, we can
determine which contribution is dominant.

Let us start with Dirac equation
\begin{equation}
\label{dirac1}
\left[\gamma^{\mu} (\partial_{\mu} - \Gamma_{\mu}) + \mu \right] \psi = 0
\end{equation}
where $\mu$ is a mass of the fermion and $\Gamma_{\mu}$ is a spin-affine
connection. Since we are interested in the Dirac fermion localized on the 
brane, we assume $\psi$ is function of $t$, $r$, $\theta = \theta_1$ and 
$\phi$. The gamma matrices in this background can 
be easily chosen as $\gamma^t = \gamma^0 / \sqrt{h}$, 
$\gamma^r = \sqrt{h} \gamma^3$, $\gamma^{\theta} = \gamma^1 / r$, and 
$\gamma^{\phi} = \gamma^2 / r \sin \theta$ where $\gamma^i$ are the 
ordinary flat-space Dirac matrices defined
\begin{equation}
\label{gammamatrix}
\gamma^0 = -i \left( \begin{array}{cc}
                     1 & 0 \\
                     0 & -1
                     \end{array}      \right) 
\hspace{2.0cm}
\gamma^i = -i \left( \begin{array}{cc}
                     0 & \sigma^i  \\
                     -\sigma^i & 0
                     \end{array}       \right)
\end{equation}
and $\sigma^i$ are usual Pauli matrices.

The spin-affine connection in the $(4+n)$-dimensional Schwarzschild background
can be straightforwardly computed with use of the affine 
connections and the results are 
\begin{eqnarray}
\label{spin-affine}
\Gamma_t&=&- \frac{(1+n) (1 - h)}{4 r} \gamma^3 \gamma^0
\hspace{2.0cm}
\Gamma_r = 0
                                \\   \nonumber
\Gamma_{\theta}&=& \frac{\sqrt{h}}{2} \gamma^3 \gamma^1
\hspace{2.0cm}
\Gamma_{\phi} = \frac{\sqrt{h}}{2} \sin \theta \gamma^3 \gamma^2 
+ \frac{1}{2} \cos \theta \gamma^1 \gamma^2.
\end{eqnarray}
Then, Dirac operator $\gamma^{\mu} (\partial_{\mu} - \Gamma_{\mu}) + \mu$
reduces to 
\begin{equation}
\label{dirac-operator}
\gamma^{\mu} (\partial_{\mu} - \Gamma_{\mu}) + \mu
= -i \left( \begin{array}{cc}
        \frac{1}{\sqrt{h}} \partial_t + i \mu & \hat{H}   \\
        -\hat{H} & -\frac{1}{\sqrt{h}} \partial_t + i \mu 
          \end{array}      \right)
\end{equation}
where 
\begin{equation}
\label{bozo31}
\hat{H} = \sqrt{h} \sigma^3 D_r + \frac{\sigma^1}{r} D_{\theta}
         + \frac{\sigma^2}{r \sin \theta} \partial_{\phi}
\end{equation}
and 
\begin{eqnarray}
\label{bozo36}
D_r&=&\partial_r + \frac{1}{r} + \frac{(n+1) (1 - h)}{4 h r}
                                            \\   \nonumber
D_{\theta}&=&\partial_{\theta}  + \frac{\cot \theta}{2}.
\end{eqnarray}

In order to perform the separation of variables we take an following
ansatz
\begin{equation}
\label{ansatz1}
\psi = \frac{e^{-i \epsilon t}}{r h^{\frac{1}{4}}}
\left( \begin{array}{c}
       G(r) \hat{\Theta}(\theta, \phi) \\
       -i F(r) \sigma^3 \hat{\Theta}(\theta, \phi)
       \end{array}     \right)
\end{equation}
where $\hat{\Theta}(\theta, \phi)$ is angle-dependent two-component spinor. 
Then, it is straightforward to derive two radial equations 
\begin{eqnarray}
\label{df-radial1}
\sqrt{h} \frac{d G}{d r}&+&\frac{k}{r} G = \left(\frac{\epsilon}{\sqrt{h}} + 
      \mu \right) F         \\   \nonumber
\sqrt{h} \frac{d F}{d r}&-&\frac{k}{r} F = \left(\frac{-\epsilon}{\sqrt{h}} + 
      \mu \right)G
\end{eqnarray}
where $k$ is defined from the angular equation
\begin{equation}
\label{angular1}
\left(\sigma^2 D_{\theta} - \frac{\sigma^1}{\sin \theta} \partial_{\phi}
           \right) \hat{\Theta} = - i k \hat{\Theta}.
\end{equation}
The angular equation was discussed long ago by Schr\"{o}dinger\cite{schro} 
and the constant $k$ is related to the orbital angular momentum $\ell$ 
by $\ell = |k + 1/2| - 1/2$ and to the total angular momentum $j$ by 
$j = |k| - 1/2$\cite{unruh76}. Thus, for example, the lowest quantum number 
$j = 1/2$
corresponds to $k = -1$ ($\ell = 0$) and $k = 1$ ($\ell = 1$). 

Removing the low component $F(r)$ in Eq.(\ref{df-radial1}), we can finally 
derive a second-order radial differential equation:
\begin{equation}
\label{df-radial2}
\frac{d^2 G}{d x^2} + 
\left[ \epsilon^2 \left(\frac{1 - \lambda \sqrt{h}}{1 + \lambda \sqrt{h}}
                           \right) - 
      \frac{k^2 h}{(1 + \lambda \sqrt{h})^2 r^2} + 
      \frac{d}{d x} 
      \left(\frac{k \sqrt{h}}{(1 + \lambda \sqrt{h}) r} \right) \right] G = 0
\end{equation}
where $\lambda = \mu / \epsilon$ and $x$ is defined as
\begin{equation}
\label{x-def1}
\frac{d x}{d r} = \frac{1 + \lambda \sqrt{h}}{h}.
\end{equation}
It is important to note that the new variable $x$ goes to 
$x \sim (r_H / (n+1)) \ln h$ in the near-horizon region and 
$x \sim (1 + \lambda) r$ in the asymptotic region. 

Due to the $\sqrt{h}$ in Eq.(\ref{df-radial2}), the method of Maldacena and 
Strominger in Ref.\cite{mal96-1} does not work in this case. Thus, we will
adopt the method Unruh did in Ref.\cite{unruh76}.

Eq.(\ref{df-radial2}) indicates that the radial equation in the near-horizon
becomes simply
\begin{equation}
\label{df-horizon1}
\frac{d^2 G_{NH}}{d x^2} + \epsilon^2 G_{NH} = 0
\end{equation}
which makes the following outgoing wave
\begin{equation}
\label{df-outgo1}
G_{NH} = \alpha_{I} e^{-i \epsilon x}
\approx \alpha_{I} e^{-i \frac{\epsilon r_H}{n+1} \ln h}.
\end{equation}

In the intermediate region where $\epsilon^2$ and $\mu^2$ are much smaller
than the other terms the radial equation (\ref{df-radial2}) reduces to 
\begin{equation}
\label{df-inter1}
\frac{d^2 G_{IM}}{d x^2} + 
\left[ \frac{-k^2 h}{(1 + \lambda \sqrt{h})^2 r^2} + 
       \frac{d}{d x} 
       \left( \frac{k \sqrt{h}}{(1 + \lambda \sqrt{h}) r} \right) \right]
        G_{IM} = 0.
\end{equation}
Defining 
\begin{equation}
\label{df-inter2}
H_{IM} \equiv \frac{d G_{IM}}{d x} + \frac{k \sqrt{h}}
                                          {(1 + \lambda \sqrt{h}) r} G_{IM},
\end{equation}
we can change Eq.(\ref{df-inter1}) into the first-order differential
equation in the form
\begin{equation}
\label{df-inter3}
\frac{d H_{IM}}{d r} - 
\frac{k}{\sqrt{h} r} H_{IM} = 0.
\end{equation}
The solution of (\ref{df-inter3}) is 
\begin{equation}
\label{df-inter4}
H_{IM} = \beta_{II} \left( \frac{1 - \sqrt{h}}{1 + \sqrt{h}} 
                    \right)^{-\frac{k}{n+1}}.
\end{equation}
Therefore, inserting Eq.(\ref{df-inter4}) into Eq.(\ref{df-inter2}), we 
can obtain the following $G_{IM}$ 
\begin{equation}
\label{df-inter5}
G_{IM} = \alpha_{II} \left( \frac{1 - \sqrt{h}}{1 + \sqrt{h}} 
                     \right)^{\frac{k}{n+1}} + 
                     \beta_{II} \cal{G}_{IM}
\end{equation}
where $\cal{G}_{IM}$ is a particular solution to 
\begin{equation}
\label{df-inter6}
\frac{d \cal{G}_{IM}}{d r} + \frac{k}{\sqrt{h} r} {\cal{G}_{IM}}
= \frac{1 + \lambda \sqrt{h}}{h} \left( \frac{1 - \sqrt{h}}{1 + \sqrt{h}} 
                     \right)^{-\frac{k}{n+1}}.
\end{equation}
For $k < 0$ we obtain
\begin{equation}
\label{parti2}
{\cal{G}_{IM}} = r_{H} \left( \frac{1 - \sqrt{h}}{1 + \sqrt{h}} 
                     \right)^{\frac{k}{n+1}}
                 \int_1^{\sqrt{h}}
                \frac{2(1 + \lambda \rho)}{(n+1) \rho}
                \frac{(1 - \rho)^{\frac{2 |k| - (n+2)}{n+1}}}
                     {(1 + \rho)^{\frac{2 |k| + (n+2)}{n+1}}} d \rho,
\end{equation}
whereas for $k > 0$ we can show
\begin{equation}
\label{parti3}
{\cal{G}_{IM}} = r_H \left( \frac{1 - \sqrt{h}}{1 + \sqrt{h}} 
                     \right)^{\frac{k}{n+1}}
                 \left\{ \frac{1}{n+1} \ln h + 
                        \int_0^{\sqrt{h}} d \rho 
                        \frac{2}{(n+1) \rho}
                        \left[\frac{(1 + \lambda \rho)}
                                    {(1 - \rho^2)^{\frac{n+2}{n+1}}}
                  \left( \frac{1 + \rho}{1 - \rho} \right)^{\frac{2 k}{n+1}}
                  -1  \right]      \right\}.
\end{equation}

In the asymptotic region Eq.(\ref{df-radial2}) reduces to 
\begin{equation}
\label{df-ff1}
\frac{d^2 G_{FF}}{d x^2} + 
\left[\epsilon^2 \frac{1 - \lambda}{1 + \lambda} - \frac{k(k+1)}{x^2}
           \right] G_{FF} = 0.
\end{equation}

Since $x \sim (1 + \lambda) r$ in the asymptotic region, the solution of 
Eq.(\ref{df-ff1}) can be written as 
\begin{equation}
\label{df-ff2}
G_{FF}(r) = \alpha_{III} \sqrt{\frac{r}{r_H}} J_{|k + \frac{1}{2}|}
                                             (\epsilon v r)
        +   \beta_{III} \sqrt{\frac{r}{r_H}} Y_{|k + \frac{1}{2}|}
                                             (\epsilon v r)
\end{equation}
where $v = \sqrt{1 - \mu^2 / \epsilon^2}$.

Now, let us consider a matching between $G_{NH}$ and $G_{IM}$. Firstly, we
note that the near-horizon solution (\ref{df-outgo1}) can be expanded
as 
\begin{equation}
\label{df-match1}
G_{NH} \sim \alpha_{I} 
\left(1 - i \frac{\epsilon r_H}{n+1} \ln h + \cdots \right).
\end{equation}
Secondly, let us consider $r \rightarrow r_H$ limit of $G_{IM}$. 
Eq.(\ref{df-inter6}) indicates the $r \rightarrow r_H$ limit of $G_{IM}$
becomes
\begin{equation}
\label{df-match2}
\lim_{r \rightarrow r_H} G_{IM} \approx \alpha_{II} + 
\beta_{II} \lim_{r \rightarrow r_H} {\cal{G}_{IM}}.
\end{equation}
For $k < 0$ Eq.(\ref{parti2}) implies
\begin{equation}
\label{df-match3}
\lim_{r \rightarrow r_H} {\cal{G}_{IM}} \sim r_H b_n + \frac{r_H}{n+1} \ln h
\end{equation}
where $b_n$ is a $n$-dependent finite quantity defined
\begin{equation}
\label{df-bn1}
b_n \sim \int_0^1 d\rho \frac{2}{(n+1) \rho}
\left[1 - (1 + \lambda \rho)
     \frac{(1 - \rho)^{\frac{2|k| - (n+2)}{n+1}}}
          {(1 + \rho)^{\frac{2|k| + (n+2)}{n+1}}}    \right].
\end{equation}
Note that the factor $1$ in the bracket compensates an infinity arising due to
$1/\rho$ at $\rho \sim 0$. Thus, comparing Eq.(\ref{df-match2}) with 
Eq.(\ref{df-match1}) simply yield the matching conditions between 
$G_{NH}$ and $G_{IM}$
\begin{equation}
\label{df-match4}
\alpha_{II} = \alpha_{I} \hspace{3.0cm}
\beta_{II} = -i \epsilon \alpha_{I}.
\end{equation}

For $k > 0$, Eq.(\ref{parti3}) implies
\begin{equation}
\label{df-match5}
\lim_{r \rightarrow r_H} G_{IM} \sim \alpha_{II} + 
\frac{\beta_{II}}{n+1} \ln h
\end{equation}
which also gives Eq.(\ref{df-match4}).

Next, let us consider a matching between $G_{IM}$ and $G_{FF}$. Firstly,
let us consider $k > 0$ case. Taking $r \rightarrow \infty$ limit to 
$G_{IM}$ in Eq.(\ref{df-inter5}) and direct integration in 
Eq.(\ref{parti3}) yields 
\begin{eqnarray}
\label{df-match6}
 \lim_{r \rightarrow \infty} G_{IM}&\sim& \alpha_{II} 4^{-\frac{k}{n+1}}
\left(\frac{r_H}{r}\right)^k + \beta_{II} 4^{-\frac{k}{n+1}}
\left(\frac{r_H}{r}\right)^k
\left[ {\cal{E}} + \frac{1 + \lambda}{2 k + 1} 4^{\frac{2 k}{n+1}}
       \left( \frac{r}{r_H} \right)^{2k+1} \right] r_H
                                                    \\   \nonumber
& & \sim \alpha_{II} 4^{-\frac{k}{n+1}} \left(\frac{r_H}{r}\right)^k 
+ \beta_{II}
\frac{(1 + \lambda) 4^{\frac{k}{n+1}}}
     {2 k + 1}
\left(\frac{r}{r_H} \right)^{k+1} r_H
\end{eqnarray}
where ${\cal{E}}$ is an integration constant assumed small with respect to
$r^{2k+1}$. 

Now, we expand $G_{FF}(r)$ in Eq.(\ref{df-ff2}) with assuming 
$\epsilon v r <<1$, which is valid in the low-energy approximation. Then, 
the usual Bessel function properties yield 
\begin{equation}
\label{df-match7}
\lim_{\epsilon v r \rightarrow 0} G_{FF}(r) \sim
\frac{\alpha_{III}}{\sqrt{r_H}}
\frac{\left(\frac{\epsilon v}{2}\right)^{k + \frac{1}{2}}}
     {\Gamma \left(k + \frac{3}{2} \right)} r^{k+1}
- \frac{\beta_{III}}{\pi \sqrt{r_H}}
  \Gamma \left(k + \frac{1}{2} \right)
  \left( \frac{\epsilon v}{2} \right)^{-\left(k + \frac{1}{2} \right)} r^{-k}.
\end{equation}
Comparision of Eq.(\ref{df-match7}) with Eq.(\ref{df-match6}), therefore, 
makes the following matching conditions
\begin{eqnarray}
\label{df-match8}
\alpha_{II}&=& - 
\frac{2^{\frac{2 k}{n+1}} \Gamma \left(k + \frac{1}{2} \right)}{\pi}
    \left( \frac{\epsilon v r_H}{2} \right)^{-\left( k + \frac{1}{2} \right)}
    \beta_{III}
                           \\  \nonumber
\beta_{II}&=&
\frac{2^{1 - \frac{2 k}{n+1}}}
     {r_H (1 + \lambda) \Gamma \left(k + \frac{1}{2} \right)}
     \left( \frac{\epsilon v r_H}{2} \right)^{ k + \frac{1}{2} }
      \alpha_{III}.
\end{eqnarray}

A similar procedure leads the following matching conditions for $k < 0$;
\begin{eqnarray}
\label{df-match9}
\alpha_{II}&=& \frac{2^{\frac{2 k}{n+1}}}
                    {\Gamma \left(\frac{1}{2} - k \right)}
   \left( \frac{\epsilon v r_H}{2} \right)^{-\left( k + \frac{1}{2} \right)}
    \alpha_{III}
                                  \\   \nonumber
\beta_{II}&=& \frac{2^{1 - \frac{2 k}{n+1}} \Gamma \left(-k + \frac{1}{2} 
                                                   \right) }
                   {\pi r_H (1 + \lambda)}
        \left( \frac{\epsilon v r_H}{2} \right)^{ k + \frac{1}{2} }
       \beta_{III}.
\end{eqnarray}           
Thus, Eq.(\ref{df-match4}) and Eq.(\ref{df-match8}) gives for $k > 0$ 
\begin{equation}
\label{df-match10}
\frac{\beta_{III}}{\alpha_{III}} = 
\frac{\pi v 2^{-\frac{4 k}{n+1}}}
     {i (1 + \lambda) \Gamma^2 \left(k + \frac{1}{2} \right)}
     \left( \frac{\epsilon v r_H}{2} \right)^{2 k}
\end{equation}
and Eq.(\ref{df-match4}) and Eq.(\ref{df-match9}) yield for $k < 0$
\begin{equation}
\label{df-match11}
\frac{\beta_{III}}{\alpha_{III}} = 
\frac{-i \pi (1 + \lambda)}
     {v 2^{-\frac{4 k}{n+1}} \Gamma^2 \left( \frac{1}{2} - k \right)}
     \left( \frac{\epsilon v r_H}{2} \right)^{-2 k}. 
\end{equation}

Now, let us compute the low-energy absorption cross section. Computing the 
flux in the asymptotic region, one can easily express the greybody factor
for Dirac fermion as 
\begin{equation}
\label{df-cross1}
\Gamma_j (\epsilon) = 
\frac{2 i \left(\frac{\beta_{III}}{\alpha_{III}} - 
                \frac{\beta_{III}^{\ast}}{\alpha_{III}^{\ast}} \right)}
     {\Bigg|1 + i \frac{\beta_{III}}{\alpha_{III}} \Bigg|^2}.
\end{equation}
Thus for $k > 0$ $\Gamma_j (\epsilon)$ reduces to 
\begin{equation}
\label{df-cross2}
\Gamma_{j, k>0} (\epsilon) = 
\frac{\left( \frac{4 \pi v 2^{-\frac{4 k}{n+1}} 
                   \left( \frac{\epsilon v r_H}{2} \right)^{2 k} }
                  {(1 + \lambda) \Gamma^2 \left(k + \frac{1}{2} \right)}
                                                     \right) }
     { \left[1 + 
             \frac{\pi v 2^{-\frac{4 k}{n+1}} \left( \frac{\epsilon v r_H}{2}
                                              \right)^{2 k}}
                  {(1 + \lambda) \Gamma^2 \left(k + \frac{1}{2} \right)}
        \right]^2}
\approx 
\frac{4 \pi v 2^{-\frac{4 k}{n+1}} 
     \left( \frac{\epsilon v r_H}{2} \right)^{2 k}}
     {(1 + \lambda) \Gamma^2 \left(k + \frac{1}{2} \right)}
\end{equation}
and for $k < 0$
\begin{equation}
\label{df-cross3}
\Gamma_{j, k<0} (\epsilon) =
\frac{  \left( \frac{4 \pi (1 + \lambda)}
                    {v 2^{-\frac{4 k}{n+1}} 
                 \Gamma^2 \left(\frac{1}{2} - k \right)
             \left( \frac{\epsilon v r_H}{2} \right)^{2 k}} \right)}
      {\left[1 + \frac{\pi (1 + \lambda)}
                      {v 2^{-\frac{4 k}{n+1}} 
                       \Gamma^2 \left(\frac{1}{2} - k \right)
                      \left( \frac{\epsilon v r_H}{2} \right)^{2 k}}
                                                  \right]^2}
\approx
\frac{4 \pi (1 + \lambda) 
      \left( \frac{\epsilon v r_H}{2} \right)^{- 2 k}} 
     {v 2^{-\frac{4 k}{n+1}}
      \Gamma^2 \left(\frac{1}{2} - k \right)}.
\end{equation}

At this stage it is worthwhile noting the following. If one introduces a spin
by Newman-Penrose formalism, only the greybody factor for each $j$-level
can be calculated, which is just sum of $\Gamma_{j, k>0} (\epsilon)$ and 
$\Gamma_{j, k<0} (\epsilon)$. Since, however, we introduced a spin by usual 
Dirac form, we are able to calculate $\Gamma_{j, k>0} (\epsilon)$ and
$\Gamma_{j, k<0} (\epsilon)$ individually, where the former corresponds to 
$\ell = j + 1/2$ and the latter to $\ell = j - 1/2$. Thus we can determine
which contribution of angular momentum quantum number is dominant. If, 
for example, $j$ is fixed, there are contribution to the greybody factor
from $\ell = j + 1/2$ (or $k = \ell$) and 
$\ell = j - 1/2$ (or $k = -\ell - 1$). Then Eq.(\ref{df-cross2}) and 
(\ref{df-cross3}) indicate 
\begin{equation}
\label{add-df-1}
\frac{\Gamma_{j, k = j + \frac{1}{2}}}
     {\Gamma_{j, k = -j - \frac{1}{2}}} = 
\frac{1 - \lambda}{1 + \lambda} \leq 1
\end{equation}
Thus the contribution from $\ell = j - 1/2$ to the greybody factor is 
larger than that from $\ell = j + 1/2$. However, in the massless limit two 
contributions are exactly same.

The absorption cross section for fixed $j$ is given by
\begin{eqnarray}
\label{df-cross4}
\sigma_j(\epsilon)&=& \frac{\pi (2 j + 1)}{2 \epsilon^2 v^2}
\left(\Gamma_{k = j + \frac{1}{2}} (\epsilon) + 
      \Gamma_{k = -j - \frac{1}{2}} (\epsilon) \right)
                                                      \\  \nonumber
&=& \frac{\pi (2 j + 1)}{\epsilon^2 v^3}
    \frac{(2 \pi) 2^{-\frac{4 j + 2}{n + 1}}}
         {2^{2 j} \Gamma^2 \left(j + 1\right)}
    (\epsilon v r_H)^{2 j + 1}.
\end{eqnarray}     

Now, let us consider the limiting cases. In the massless limit 
Eq.(\ref{df-cross4}) shows
\begin{equation}
\label{df-cross5}
\sigma_j^{m=0}(\epsilon) = \frac{\pi (2 j + 1)}{\epsilon^2}
\frac{(2 \pi) 2^{-\frac{4 j + 2}{n + 1}}}
     {2^{2 j} \Gamma^2 \left(j + 1\right)}
(\epsilon r_H)^{2 j + 1}
\end{equation}
which exactly coincides with Eq.(45) of Ref.\cite{kanti03-1}. 

Next, let us 
consider the case of the lowest angular momentum quantum number,
{\it i.e.} $j = 1/2$. In this case the absorption cross section becomes 
\begin{equation}
\label{df-cross6}
\sigma_{j=\frac{1}{2}} (\epsilon) = \frac{\pi}{v}
2^{\frac{3n-1}{n+1}} r_H^2.
\end{equation}
Since Eq.(\ref{nomass1}) indicates the low-energy absorption cross section 
for S-wave is 
\begin{equation}
\label{df-cross7}
\sigma_{\ell = 0} = \frac{4 \pi r_H^2}{v},
\end{equation}
the ratio between $\sigma_{j = 1/2}$ and $\sigma_{\ell = 0}$ when 
$\epsilon = \omega$ and $\mu = m$ becomes
\begin{equation}
\label{ratio1}
\frac{\sigma_{j=\frac{1}{2}}}{\sigma_{\ell = 0}} = 
2^{\frac{n-3}{n+1}}.
\end{equation}
Thus we get $\sigma_{j = 1/2} / \sigma_{\ell = 0} = 1/8$ when $n=0$, which 
Unruh found in his seminar paper in Ref.\cite{unruh76}. Thus, our result 
Eq.(\ref{ratio1}) is a generalized ratio between the low-energy
absorption cross sections for spin-$1/2$ and scalar particles in the 
higher-dimensional brane-world theories. In conclusion, our result
(\ref{df-cross4}) correctly reproduces the $n=0$ limit as well as 
$4d$ limit.

\section{Low-Energy Absorption Cross Section for Bulk Scalar}
In this section we will discuss on the low-energy absorption problem 
for the minimally-coupled massive scalar which lives in the bulk. Thus, we
should assume the scalar field $\Phi$ is function of all bulk coordimates.
Then, it is straightforward to show that the usual Klein-Gordon equation
$(\Box - m^2) \Phi = 0$ in the $(4+n)$-dimensional Schwarzschild background
(\ref{schwarz1}) reduces to
\begin{equation}
\label{radial4-1}
\frac{h(r)}{r^{n+2}} \frac{d}{d r}
\left[h(r) r^{n+2} \frac{d R}{d r} \right] + 
\left[ \omega^2 - h(r) \left\{m^2 + \frac{\ell (\ell + n + 1)}{r^2} \right\}
                                                 \right] R = 0
\end{equation}
where we used 
$\Phi = e^{-i \omega t} R_{\omega \ell}(r) \tilde{Y}_{\ell}(\Omega)$ and
$\tilde{Y}_{\ell}(\Omega)$ is an higher-dimensional spherical harmonics.

At this stage it is worthwhile noting that one can calculate the low-energy
absorption cross section by using the method used in Ref.\cite{mal96-1}
and Unruh's original method used in Ref.\cite{unruh76}. Since both 
procedures yield a same result, we will adopt the latter in this paper.

In order to obtain the near-horizon solution of Eq.(\ref{radial4-1}) it is 
convenient to introduce a tortoise coordinate
\begin{equation}
\label{tortoise4-1}
y = \frac{1}{(n+1) r_H^{n+1}} \ln h(r)
\end{equation}
which changes Eq.(\ref{radial4-1}) into 
\begin{equation}
\label{radial4-2}
\frac{d^2 R}{d y^2} + r^{2 n + 4} 
\left[ \omega^2 - h(r) \left\{ m^2 + \frac{\ell (\ell + n + 1)}{r^2} \right\}
                                           \right]R = 0.
\end{equation}
Since $h \sim 0$ in the near-horizon region, Eq.(\ref{radial4-2}) 
approximately reduces to 
\begin{equation}
\label{near4-1}
\frac{d^2 R_{NH}}{d y^2} + r_H^{2n+4} \omega^2 R_{NH} \approx 0
\end{equation}
which gives the out-going wave
\begin{equation}
\label{near4-2}
R_{NH} = A_I e^{-i \omega r_H^{n+2} y}.
\end{equation}

In the intermediate region we use an inequality
$h(r) \ell (\ell + n + 1) / r^2 >> \omega^2$ which makes Eq.(\ref{radial4-1})
to be in this region
\begin{equation}
\label{inter4-1}
\frac{d}{d r}
\left[ h(r) r^{n+2} \frac{d R_{IM}}{d r} \right] - r^n \ell (\ell + n + 1)
R_{IM} = 0.
\end{equation}
It is easy to show that Eq.(\ref{inter4-1}) provides a solution in terms of 
the Legendre polynomials as following
\begin{equation}
\label{inter4-2}
R_{IM} = A_{II} P_{\frac{\ell}{n+1}} \left( 2 \left(\frac{r}{r_H} \right)^{n+1}
                                            -1 \right)
        + B_{II} Q_{\frac{\ell}{n+1}} \left( 2 \left(\frac{r}{r_H} \right)^{n+1}
                                            -1 \right).
\end{equation}

In the asymptotic region $h(r) \sim 1$ and the radial equation 
(\ref{radial4-1}) becomes approximately 
\begin{equation}
\label{asym4-1}
\frac{1}{r^{n+2}} \frac{d}{d r}
r^{n+2} \frac{d R_{FF}}{d r} + 
\left[ (\omega^2 - m^2) - \frac{\ell (\ell + n + 1)}{r^2} \right] R_{FF} = 0.
\end{equation}
It is easy to show that Eq.(\ref{asym4-1}) is solved by
\begin{equation}
\label{asym4-2}
R_{FF} = \frac{1}{r^{\frac{n+1}{2}}}
\left[ A_{III} J_{\ell + \frac{n+1}{2}} (\omega v r) + 
       B_{III} Y_{\ell + \frac{n+1}{2}} (\omega v r) \right]
\end{equation}
where $v = \sqrt{1 - m^2 / \omega^2}$. 

Now, let us consider a matching between $R_{NH}$ and $R_{IM}$. Using a 
relations between the Legendre polynomials and the hypergeometric finction
\begin{eqnarray}
\label{bozo4-1}
P_{\nu}(z)&=& F\left(-\nu, \nu+1; 1; \frac{1 - z}{2} \right)
                                                     \\ \nonumber
Q_{\nu}(z)&=& 2^{-\nu - 1} \sqrt{\pi}
\frac{\Gamma(\nu + 1)}{\Gamma \left(\nu + \frac{3}{2}\right)}
z^{-\nu - 1} F\left(1 + \frac{\nu}{2}, \frac{1}{2} + \frac{\nu}{2};
                   \nu + \frac{3}{2}; \frac{1}{z^2} \right),
\end{eqnarray}
the $r \rightarrow r_H$ limit of $R_{IM}$ reduces to 
\begin{eqnarray}
\label{match4-1}
\lim_{r \rightarrow r_H} R_{IM}&=&
A_{II} \left[1 + \ell \left( \frac{\ell}{n+1} + 1 \right)
\left( \frac{r}{r_H} - 1\right) + \cdots \right]
                                                  \\  \nonumber
&+& \frac{B_{II}}{2}
\Bigg[ -\ln \left( \frac{r}{r_H} - 1\right) + 
\Bigg\{ 2 \psi(1) - \psi \left(1 + \frac{\ell}{2 n + 2}  \right) - 
\psi \left( \frac{1}{2} + \frac{\ell}{2 n + 2} \right) 
                                                 \\  \nonumber
& & \hspace{9.0cm} - \ln 4 (n+1) \Bigg\} 
+ \cdots \Bigg]
\end{eqnarray}
where $\psi(z)$ is an usual digamma function. Comparing Eq.(\ref{match4-1}) 
with 
\begin{equation}
\label{match4-2}
R_{NH} \sim A_I \left[ 1 - \frac{i r_H \omega}{n+1} 
\left\{ \ln \left( \frac{r}{r_H} - 1\right) + \ln (n+1) \right\} + \cdots
                                          \right],
\end{equation}
we can derive the matching conditions
\begin{eqnarray}
\label{match4-3}
A_{II}&=&A_I \left[1 + \frac{i r_H \omega}{n+1}
\left\{\psi \left(1 + \frac{\ell}{2 n + 2}  \right) + 
       \psi \left( \frac{1}{2} + \frac{\ell}{2 n + 2} \right) + \gamma
     - \psi \left(\frac{1}{2}\right) \right\}  \right] \approx A_I
                                                  \\   \nonumber
B_{II}&=& \frac{2 i r_H \omega}{n+1} A_I
\end{eqnarray}
where $\gamma$ is an Euler's constant and the last approximation in 
$A_{II}$ comes from the low-energy approximation.

Next the matching between $R_{IM}$ and $R_{FF}$ will be discussed. Using  
relations (\ref{bozo4-1}) it is not difficult to show that the 
$r \rightarrow \infty$ limit of $R_{IM}$ is 
\begin{equation}
\label{match4-4}
\lim_{r \rightarrow \infty} R_{IM} = A_{II}
\frac{\Gamma \left(1 + \frac{2 \ell}{n+1} \right)}
     {\Gamma^2 \left(1 + \frac{\ell}{n+1} \right)}
     \left(\frac{r}{r_H}\right)^{\ell}
+ B_{II} \frac{\Gamma^2 \left(1 + \frac{\ell}{n+1} \right)}
              {2 \Gamma \left(2 + \frac{2 \ell}{n+1} \right)}
     \left(\frac{r_H}{r}\right)^{n+ \ell + 1}.
\end{equation}
Since $\omega v r \rightarrow 0$ limit of $R_{FF}$ is 
\begin{equation}
\label{match4-5}
\lim_{\omega v r \rightarrow 0} R_{FF} = 
A_{III} \frac{\left(\frac{\omega v}{2} \right)^{\ell + \frac{n+1}{2}}}
             {\Gamma \left( \ell + \frac{n+3}{2} \right)} r^{\ell} - 
\frac{B_{III}}{\pi} \Gamma \left(\ell + \frac{n+1}{2} \right)
\left( \frac{2}{\omega v} \right)^{\ell + \frac{n+1}{2}}
r^{-\ell - n - 1},
\end{equation}
the comparision of Eq.(\ref{match4-4}) with Eq.(\ref{match4-5}) gives the 
following natching conditions;
\begin{eqnarray}
\label{match4-6}
A_{III}&=& \frac{\Gamma \left(1 + \frac{2 \ell}{n+1} \right)
                 \Gamma \left(\ell + \frac{n+3}{2} \right)
                }
                {r_H^{\ell} \Gamma^2 \left(1 + \frac{\ell}{n+1} \right)
                }
\left(\frac{\omega v}{2} \right)^{-\left(\ell + \frac{n+1}{2} \right)} A_{II}
                                       \\   \nonumber
B_{III}&=& - \frac{\pi r_H^{n + \ell + 1} \Gamma^2 
                                        \left(1 + \frac{\ell}{n+1} \right)
                  }
                  {2 \Gamma \left(2 + \frac{2 \ell}{n+1} \right)
                     \Gamma \left(\ell + \frac{n+1}{2} \right)
                  }
\left(\frac{\omega v}{2} \right)^{\ell + \frac{n+1}{2}} B_{II}.
\end{eqnarray}
One should note that the expansion (\ref{match4-5}) of $R_{FF}$ is valid in 
the low-energy limit.

Now, it is straightforward to calculate the low-energy absorption cross section
for the bulk scalar. For the computation we first separate $R_{FF}$ as a 
combination of incident and reflected waves;
\begin{equation}
\label{absorp4-1}
\lim_{r \rightarrow \infty} R_{FF} = \phi_{in} + \phi_{re}
\end{equation}
where
\begin{eqnarray}
\label{absorp4-2}
\phi_{in}&=&e^{i \frac{\pi}{2} \left(\ell + \frac{n}{2} + 1\right)}
\sqrt{\frac{1}{2 \pi \omega v}} (A_{III} + i B_{III})
\frac{e^{-i \omega v r}}{r^{\frac{n}{2} + 1}}
                                               \\   \nonumber
\phi_{re}&=&e^{-i \frac{\pi}{2} \left(\ell + \frac{n}{2} + 1\right)}
\sqrt{\frac{1}{2 \pi \omega v}} (A_{III} - i B_{III})
\frac{e^{i \omega v r}}{r^{\frac{n}{2} + 1}}.
\end{eqnarray}
Then, the greybody factor is directly computed from the reflection 
coefficient
\begin{equation}
\label{absorp4-3}
\Gamma_{\ell} (\omega) = 1 - 
\Bigg| \frac{1 - i \frac{B_{III}}{A_{III}}}
            {1 + i \frac{B_{III}}{A_{III}}} \Bigg|^2.
\end{equation}
The matching conditions (\ref{match4-3}) and (\ref{match4-6}) gives
\begin{equation}
\label{absorp4-4}
\frac{B_{III}}{A_{III}} = -i D
\end{equation}
where
\begin{equation}
\label{absorp4-5}
D = \frac{\pi^2}{2^{\frac{4 \ell}{n+1}} v} 
\left(\frac{\omega v r_H}{2} \right)^{n+ 2 \ell + 2}
\frac{\Gamma^2 \left(1 + \frac{\ell}{n+1} \right)}
     {\Gamma^2 \left( \frac{1}{2} + \frac{\ell}{n+1} \right)
      \Gamma^2 \left(\frac{n+3}{2} + \ell \right)}.
\end{equation}
Thus, the factor $\Gamma_{\ell}(\omega)$ reduces to 
\begin{equation}
\label{absorp4-6}
\Gamma_{\ell}(\omega) = \frac{4 D}{(1 + D)^2}
\approx 4 D = 
\frac{ 4 \pi^2}{2^{\frac{4 \ell}{n+1}} v} 
\left(\frac{\omega v r_H}{2} \right)^{n+ 2 \ell + 2}
\frac{\Gamma^2 \left(1 + \frac{\ell}{n+1} \right)}
     {\Gamma^2 \left( \frac{1}{2} + \frac{\ell}{n+1} \right)
      \Gamma^2 \left(\frac{n+3}{2} + \ell \right)}.
\end{equation}
The massless limit of Eq.(\ref{absorp4-6}) exactly coincides with 
Eq.(37) of Ref.\cite{kanti02-1}.

The relation
between the absorption cross section and the greybody factor can be 
derived using the $(4+n)$-dimensional optical theorem \cite{gub97-1}
\begin{equation}
\label{optical}
\sigma_{\ell} (\omega) = 
\frac{2^n \Gamma^2 \left(\frac{n+3}{2}\right)}
     {\pi (\omega v r_H)^{n+2}}
     \tilde{A}_H
\frac{(2 \ell + n + 1) (\ell + n)!}{(n + 1)! \ell !}
\Gamma_{\ell}(\omega)
\end{equation}
where $\tilde{A}_H$ is an area of the horizon hypersurface
\begin{equation}
\label{hypers}
\tilde{A}_H = \frac{2 \pi^{\frac{n+3}{2}}}{\Gamma \left( \frac{n+3}{2} \right)}
             r_H^{n+2}.
\end{equation}

The most important quantity is a low-energy absorption cross section for 
S-wave, which is derived by letting $\ell = 0$ in Eq.(\ref{optical})
\begin{equation}
\label{final4-1}
\sigma_{\ell = 0}(\omega) = \frac{\tilde{A}_H}{v}.
\end{equation}
Thus, at the massless limit the absorption cross section for S-wave exactly
coincides with the area of the horizon hypersurface $\tilde{A}_H$, which
is an higher dimensional generalization of the universality. The $4d$ limit
of $\sigma_{\ell}$ is easily computed by letting $n = 0$, which is exactly same
with Eq.(\ref{noextra1}). Also the massless limit of Eq.(\ref{optical}) exactly
coincides with Eq.(5.4) of Ref.\cite{harris03-1}. Thus our result has correct 
$4d$ and massless limits.

\section{Conclusion}
The low-energy absorption cross section for the brane-localized massive scalar,
brane-localized massive Dirac fermion, and massive bulk scalar are explicitly
computed when the background is $(4+n)$-dimensional Schwarzschild spacetime. 

For the case of the brane-localized massive scalar our result (\ref{partial1})
has a correct massless limit. But the $4d$ limit of Eq.(\ref{partial1}) is not
exactly same with the Unruh's $4d$ result (\ref{unruh1}), 
but coincides with its 
low-energy expansion. The reason for this is clarified in the appendix. 

For the case of the brane-localized Dirac fermion we introduced the 
particle's spin using the traditional Dirac form instead of the 
Newman-Penrose formalism. Thus, we should compute the spin-affine 
connections and separate the Dirac equation to derive a radial equation, 
which were performed explicitly in section III. Our introduction of spin
enables us to compute the contributions to the low-energy absorption 
cross section from the orbital angular momentum quantum numbers $\ell$. Since, 
for spin-$1/2$ particle, total angular momentum quantum number 
$j$ is contributed from 
$\ell = j \pm 1/2$, the ratio of $\sigma_{j, \ell = j + 1/2}$ and 
$\sigma_{j, \ell = j - 1/2}$ is given in Eq.(\ref{add-df-1}). This equation
indicates that the contribution from lower orbital angular momentum is 
larger than that from higher orbital angular momentum in the massive case. 
However, two contributions are exactly same in the massless limit.
The ratio of 
$\sigma_{j = 1/2}$ and the low-energy absorption cross section for S-wave 
massive scalar turns out to be dependent on the number of extra dimensions
as $2^{(n-3)/(n+1)}$. Thus, we can reproduce the ratio factor $1/8$ at
$n=0$ which was shown by Unruh in Ref.\cite{unruh76}. Of course, our result
has a correct massless limit.

For the case of the bulk scalar our result (\ref{optical}) for the low-energy
absorption cross section is shown to have correct massless and $4d$ limits.
Especially, the cross section for the massless S-wave is exactly same with
the area of the horizon hypersurface, which is a higher-dimensional 
generalization of the universality for S-wave.

The extensions of our paper can be pursued in the several different directions.
Firstly, one may consider the low-energy absorption cross section for a massive
particle which has an arbitrary spin. In this case we should derive a radial
master equation using the Newman-Penrose formalism. Then, it may be possible to 
examine the effect of spin in the absorption 
and emission problems of the higher-dimensional
theories. Similar approach can be applied to generalize Ref.\cite{cve98-1}. 
Then we may understand the effect of particle's mass in the stringy description
of black hole.

Since the matching method between the near-horizon and asymptotic solutions 
introduced in Ref.\cite{park00,jung03} gives an information on the 
high-energy absorption cross section, it may be possible to extend our
paper to the extremely high-energy domain. It is of interest to check the 
effect of the extra dimensions in the high-energy absorption and emission
problems.

Another extension of the present paper is to examine the absorption and 
emission  problems in the entire range of energy. This can be performed 
numerically by following our recent paper \cite{jung04}. In Ref.\cite{jung04}
the absorption cross section for massive S-wave in the $4d$ black hole 
background exhibits a pecular behavior, {\it i.e.} decreasing behavior with
increasing energy in the extremely low-energy regime. This behavior, as a 
result, breaks the universality for S-wave. It is interesting to check whether
or not similar behavior exists in the higher-dimensional theories. 

The most interesting one seems to be to understand the physical
origin of the ratio factor $2^{(n-3)/(n+1)}$. Since the absorption cross 
section for the brane-localized massless S-wave is fixed at $4 \pi r_H^2$,
this ratio factor indicates that the absorption cross section for the 
Dirac fermion increases when the extra dimensions exist. Especially, for the 
infinite extra dimensions the cross section for fermion becomes maximum,
twice the cross section for scalar. Thus this ratio factor can be 
important to determine the number of the extra dimensions in the future
collider.

\vspace{1cm}

{\bf Acknowledgement}:  
This work was supported by the Korea Research Foundation
Grant (KRF-2003-015-C00109).

\newpage
\begin{appendix}{\centerline{\bf Appendix}}
\setcounter{equation}{0}
\renewcommand{\theequation}{A.\arabic{equation}}
In this appendix we would like to show why the $n \rightarrow 0$ limit of 
our result (\ref{partial1}) coincides with only the low-energy expansion
of the usual $4d$ limit (\ref{unruh1}). As commented in section II the 
multiplicative and exponential factors in Eq.(\ref{unruh1}) are results of 
the representation in terms of the Coulomb wave functions for an asymptotic
solution in $4d$ case. We will show in this appendix that the expression of 
the asymptotic solution in terms of the Coulomb wave functions is impossible 
when extra dimensions exist.

For the proof it is convenient to rewrite Eq.(\ref{radial1}) as
\begin{equation}
\label{appen1}
\frac{d^2 R}{d r^2} + \frac{1}{h r}
\left[ (n+1) (1 - h) + 2 h \right] \frac{d R}{d r} +
\left[ \frac{\omega^2}{h^2} - \frac{1}{h}
      \left\{m^2 + \frac{\ell (\ell + 1)}{r^2} \right\} \right] R = 0.
\end{equation}
Using the fact that the first two terms in Eq.(\ref{appen1}) can be 
expressed as 
$$\frac{1}{\sqrt{h} r} \frac{d^2}{dr^2}
\left(\sqrt{h} r R\right) - \mbox{(no derivative terms)},$$
we can rewrite Eq.(\ref{appen1}) in the following form;
\begin{equation}
\label{appen2}
\left[\frac{d^2}{d r^2} + 
\left\{ \frac{(n+1) (1 - h)}{4 r^2 h^2} \left[(n-1) h + (n + 1)\right]
       + \frac{\omega^2}{h^2} - \frac{1}{h} \left(m^2 + \frac{\ell (\ell + 1)}
                                                           {r^2} \right)
\right\} \right]
(\sqrt{h} r R) = 0.
\end{equation}

Now, let's taking $r \rightarrow \infty$ limit in Eq.(\ref{appen2}). Choosing 
the leading terms in the asymptotic region, we can show Eq.(\ref{appen2})
reduces to 
\begin{equation}
\label{appen3}
\left[ \frac{d^2}{d r^2} + 
      \left\{ (\omega^2 - m^2) + (2 \omega^2 - m^2) 
              \left(\frac{r_H}{r}\right)^{n+1} -
             \frac{\ell (\ell + 1)}{r^2} \right\} \right]
(r R_{FF}) = 0.
\end{equation}
For $n=0$ Eq.(\ref{appen3}) yields a following asymptotic solution 
in terms of the Coulomb wave functions as expected
\begin{equation}
\label{appen4}
R_{FF}^{n=0} = A_I \frac{F_{\ell}^C \left(-\frac{(1+v^2) \omega r_H}{2 v},
                                          \omega v r \right)} {r} +
               A_{II}
                   \frac{G_{\ell}^C \left(-\frac{(1+v^2) \omega r_H}{2 v},
                                          \omega v r \right)} {r}.
\end{equation}
However, when extra dimensions exist, the asymptotic solution is expressed in 
terms of the Bessel function. Thus, the multiplicative factor and the 
exponential factor in the denominator of the Unruh's result (\ref{unruh1}) 
seem
to be valid only when there is no extra dimension.

\end{appendix} 

\begin{figure}
\caption{Plot of effective potential $V_{eff}$ given in Eq.(\ref{bozo1}). 
We plotted the $n$-dependence of the effective potential at $\ell = 5$.
Since the barrier height increases with increasing the number of extra 
dimensions, this figure indicates the existence of the extra dimensions 
may decrease the absorption cross section for the massive scalar.} 
\end{figure}

\newpage
\epsfysize=5cm \epsfbox{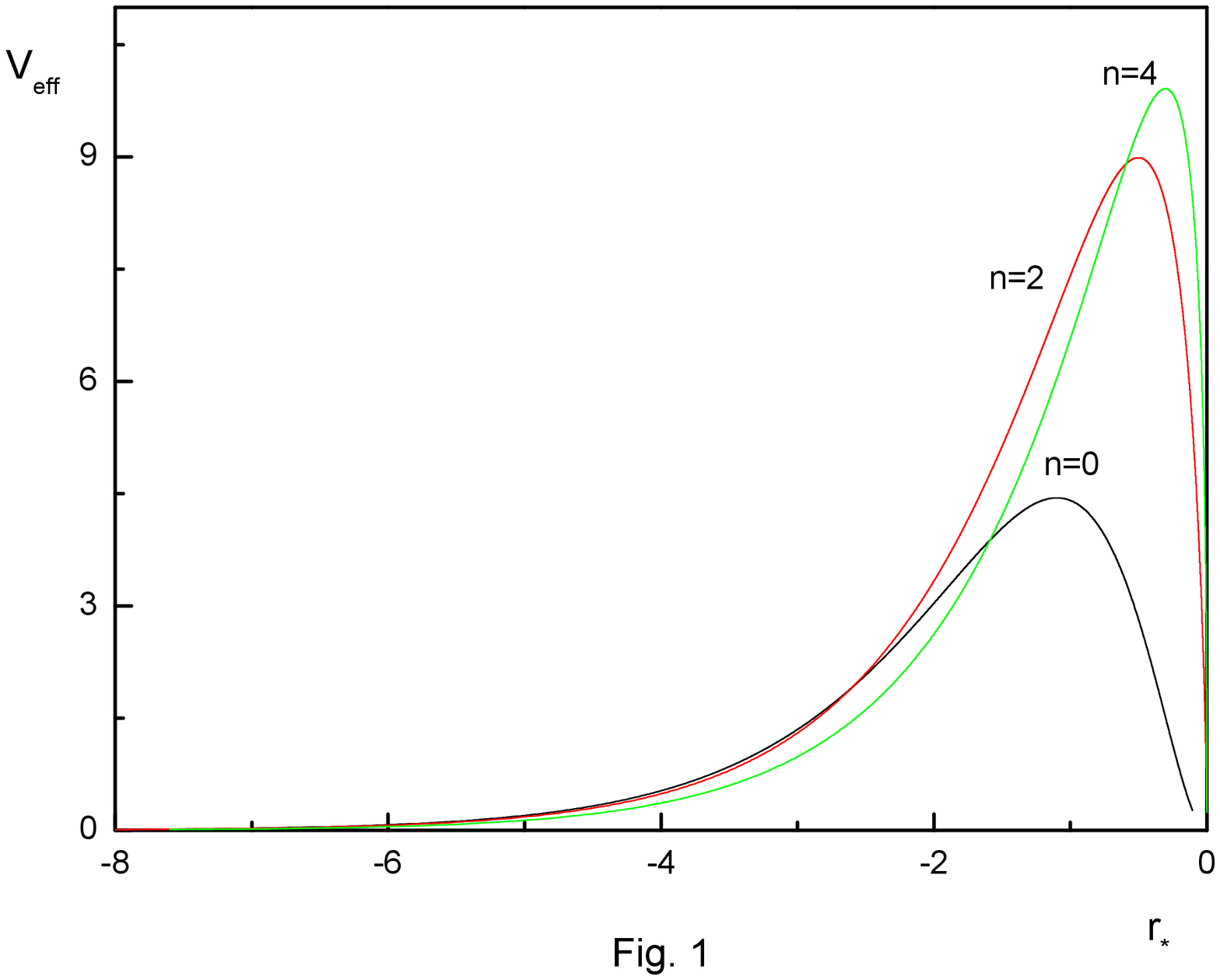}
\end{document}